\documentstyle[epsfig]{cupconf}

\ifoldfss
\else
  \ifnfssone
    \newmathalphabet{\mathit}
      \addtoversion{normal}{\mathit}{cmr}{m}{it}
      \addtoversion{bold}{\mathit}{cmr}{bx}{it}
    \newmathalphabet{\mathcal}
      \addtoversion{normal}{\mathcal}{cmsy}{m}{n}
    \else
    \ifnfsstwo
    \fi
  \fi
\fi

\def\sun{$_{\scriptscriptstyle \odot}$}
\def\sune{_{\scriptscriptstyle \odot}}

\def\ltaprx{\mathrel{\hbox{\rlap{\hbox{\lower4pt\hbox{$\sim$}}}\hbox{$<$}}}}
\def\gtaprx{\mathrel{\hbox{\rlap{\hbox{\lower4pt\hbox{$\sim$}}}\hbox{$>$}}}}

\def\upi{\pi} 
\def\umu{\mu} 
\def\hexnumber#1{\ifcase#1 0\or1\or2\or3\or4\or5\or6\or7\or8\or9\or
 A\or B\or C\or D\or E\or F\fi }

\font\eurmten=eurm10
\font\eurmseven=eurm10 at 7pt
\font\eurmfive=eurm10 at 5pt
\newfam\eurmfam
\textfont\eurmfam=\eurmten
\scriptfont\eurmfam=\eurmseven
\scriptscriptfont\eurmfam=\eurmfive
\edef\eurm@{\hexnumber\eurmfam}
 
\mathchardef\upi="0\eurm@19   
\mathchardef\umu="0\eurm@16   

\font\msxten=msam10
\font\msxseven=msam10 at 7pt
\font\msxfive=msam10 at 5pt
\newfam\msxfam
\textfont\msxfam=\msxten
\scriptfont\msxfam=\msxseven
\scriptscriptfont\msxfam=\msxfive
\edef\msx@{\hexnumber\msxfam}

\mathchardef\leqslant="3\msx@36
\mathchardef\geqslant="3\msx@3E

\makeatletter
\ifx\CUP@mtlplain@loaded\undefined
\else
\fi
\makeatother
 \makeatletter
 \ifx\CUP@mtlplain@loaded\undefined
   \font\tenbmi=cmmib10 at 10pt
   \font\sevenbmi=cmmib10 at 7pt
   \font\fivebmi=cmmib10 at 5pt

   \newfam\bmifam
   \textfont\bmifam=\tenbmi
   \scriptfont\bmifam=\sevenbmi
   \scriptscriptfont\bmifam=\fivebmi
   
 \fi
 \makeatother
\ifnfsstwo

\fi
\ifnfssone

\fi
\ifoldfss

\fi

\mathchardef\varLambda="0103

\makeatletter
\ifx\CUP@mtlplain@loaded\undefined
\else
\fi
\makeatother
\makeatletter
\ifx\CUP@mtlplain@loaded\undefined
  \font\tenbms=cmbsy10
  \font\sevenbms=cmbsy10 at 7pt
  \font\fivebms=cmbsy10 at 5pt
  \newfam\bmsfam
  \textfont\bmsfam=\tenbms
  \scriptfont\bmsfam=\sevenbms
  \scriptscriptfont\bmsfam=\fivebms

  \edef\bsy@{\hexnumber\bmsfam}
  \mathchardef\bnabla="0\bsy@72
\fi
\makeatother
\title[Collapsars, GRBs, and Supernovae]{Collapsars, Gamma-Ray Bursts,
and Supernovae}

\author[S. E. Woosley {\it et al.\/}]%
{\ns S.\ns E.\ns W\ls O\ls O\ls S\ls L\ls E\ls Y\ns\\
A.\ls I.\ns M\ls A\ls C\ls F\ls A\ls D\ls Y\ls E\ls N\ns\\
\and \ns A\ls L\ls E\ls X\ls A\ls N\ls D\ls E\ls R\ns \ns H\ls E\ls G\ls E\ls R}

\affiliation{Department of Astronomy and Astrophysics, UCSC,
Santa Cruz, CA 95064 }

\begin{document}
\ifnfssone
\else
  \ifnfsstwo
  \else
    \ifoldfss
      \let\mathcal\cal
      \let\mathrm\rm
      \let\mathsf\sf
    \fi
  \fi
\fi

\maketitle

\begin{abstract}
A diverse range of phenomena is possible when a black hole experiences
very rapid accretion from a disk due to the incomplete explosion of a
massive presupernova star endowed with rotation. In the most extreme
case, the outgoing shock fails promptly in a rotating helium star, a
black hole and an accretion disk form, and a strong gamma-ray burst
(GRB) results. However, there may also be more frequently realized
cases where the black hole forms after a delay of from several tens of
seconds to several hours as $\sim$0.1 to 5 M\sun \ falls back into
the collapsed remnant following a mildly successful supernova
explosion. There, the same MHD mechanisms frequently invoked to
produce GRBs would also produce jets in stars already in the process
of exploding. The presupernova star could be a Wolf-Rayet star or a
red or blue supergiant. Depending upon its initial pressure, the
collimation of the jet may also vary since ``hot'' jets will tend to
diverge and share their energy with the rest of the star. From these
situations, one expects diverse outcomes ranging from GRBs with a
large range of energies and durations, to asymmetric, energetic
supernovae with weak GRBs.  SN 1998bw may have been the explosion of a
star in which fall back produced a black hole and a less collimated
jet than in the case of prompt black hole formation.
\end{abstract}

\section{Introduction}
\label{intro}

In recent years, our theoretical understanding of common GRBs has
moved out of the ``dark ages'' of the 1980's into a BATSE and
Beppo-Sax inspired ``rennaisance''. The burst and its afterglow in
various wavelengths have been successfully modeled as the interaction
of a highly relativistic jet ($\gamma \gtaprx 100$) with itself
(internal shocks) and with circumstellar or interstellar material -
the so called ``relativistic fireball model'' (e.g., Piran 1999;
Meszaros 1999). The origin of the jet is still widely debated, but is
generally believed to involve the formation of a stellar mass
(approximately 2 to 5 M\sun) black hole and the rapid accretion of
matter into that hole from a disk. Modes of forming the black hole
vary (e.g., Fryer, Woosley, \& Hartmann 1999), as do assumptions
regarding the accretion rate, duration, and means of extracting disk
binding energy and converting it into the relativistic motion of the
jet. For accretion rates in excess of $\sim$0.05 M\sun \ s$^{-1}$
neutrino energy transport may be efficient (Popham, Woosley, \& Fryer
1999). For lower accretion rates, and perhaps also for the higher
ones, MHD processes - magnetic field reconnection in the disk,
extraction of black hole spin energy, Alfven waves,
magneto-centrifugal winds, etc. - are invoked.

As our understanding of GRBs has improved, an interesting ``paradigm
shift'' has also been going on in the modeling of supernovae.  For the
last 30 years, most researchers have assumed a Type II (or Ib)
supernova to be a consequence of neutrinos extracting a portion of the
binding energy of a newly formed neutron star. The neutrinos then
deposit a portion of their energy in a low density region just outside
the neutron star and the resulting ``bubble'' of pairs and radiation
explodes the rest of the star. There have been interesting exceptions
along the way (e.g., LeBlanc \& Wilson 1970; Bodenheimer \& Woosley
1983), but, for the most part, researchers have preferred their
supernovae round and without magnetic fields. 

Three things have happened lately to make us suspect that this is not
always the way supernovae work (though, admittedly, the exceptions may
be rare). First, we have observed supernovae, notably SN 1997cy and SN
1998bw, that do not fit the traditional mold (Germany et al. 1999;
Galama et al. 1998), supernovae that seem to require an order of
magnitude more energy than the traditional mechanism provides and
which may be associated with GRBs.  Second, models for GRBs have
converged on a massive presupernova star - and its explosion as a
``hypernova'' - as one leading candidate. Finally, supernovae may have
been observed as the counterparts to two or more GRBs (990425, Galama
et al. 1998; 980326, Bloom et al. 1999; 970228, Reichart 1999). Rather
suddenly, the supernova community and GRB community have awakened to
realize just how much they have in common.

In this paper, we explore this interface between GRBs and
supernovae. We find that massive stars can produce a variety of
energetic explosions ranging from traditional supernovae (by far the
most frequent occurrence) to energetic GRBs, and seemingly all points
in between.  Ordinary supernovae still come from neutron star
formation in the approximately spherically symmetric explosion of a
massive (M $\gtaprx 8$ M\sun) star with little or no fall back, but
failed or weak explosions in rotating stars give hyper-accreting black
holes whose jets can both explode the star in a grossly asymmetric way
and produce a variety of high energy phenomena.

\section{GRB Models}
\label{scenarios}

One leading model for a GRB involves a neutron star merging with
another neutron star or with a black hole. Either way, after the
merger, a black hole ends up accreting $\sim$0.01 M\sun \
(neutron star companion) to, at most, $\sim$0.5 M\sun \ (black hole
companion) from a Keplerian disk. Even for this relatively simple
model, assumptions and results vary widely. If the disk viscosity is
high, say $\alpha \gtaprx 0.01$, the disk becomes very hot and emits
its binding energy as neutrinos. Neutrino annihilation along the axis
may then energize the jet (Ruffert \& Janka 1999; Janka, Ruffert, \&
Eberl 1999; Janka et al. 1999; Rosswog et al. 1999). Since the
efficiency for neutrino annihilation is small, typically $\ltaprx$1\%,
and the viscous time scale, short ($\ltaprx$100 ms), this variety of
model produces relatively weak, brief jets, perhaps appropriate for
the class of short, hard GRBs, but unlikely to explain long energetic
events like those localized by BeppoSax. The MHD variety of this model
(e.g., Meszaros 1999) assumes a much lower viscosity and thus a
longer time scale for the accretion, up to tens of seconds. The merit
of this sort of model is that one can assume (within a large error
bar) a high efficiency for extracting energy from the disk or rotating
hole. This greater energy and longer time scale are both necessary and
sufficient to explain the most energetic bursts observed so far.

Another leading model, and the main subject of this paper, is the {\sl
collapsar}. A collapsar is a black hole formed by the incomplete
explosion of a rapidly rotating massive star (Woosley 1993; MacFadyen
\& Woosley 1999, henceforth MW99). It sets up the same sorts of
circumstances as the merging neutron star model for GRBs, but with a
number of important distinctions: 1) the event occurs only in the most
massive stars and thus tracks star formation directly; 2) a supernova
is produced by every GRB because the jet not only makes a GRB, but
explodes the star; 3) the amount of matter available for accretion
(and thus the maximum energy available for the GRB) is one to three
orders of magnitude greater than for merging compact objects; 4) the
duration of the jet is set by the collapse time scale of the star, not
by the disk viscous time scale; no very short bursts are possible; 5)
the accretion rate is likely to be lower than for the neutrino version
of merging neutron stars, but faster than some MHD versions; 6) the
engine is deeply embedded in a star that the jet must penetrate in
order to make the GRB; 7) the star is surrounded by an extended
presupernova ``wind zone'' in which the mass density is proportional
to $r^{-2}$; and 8) compared to merging neutron stars, the
gravitational radiation accompanying the burst is very weak. The
angular momentum one invokes in the collapsar model is also much less
certain than for compact objects merging by gravitational
radiation. Once the disk is set up, however, the same physics that
makes jets in merging neutron star models, be it neutrinos or MHD,
should work equally well for collapsars. The interaction of this jet
with the rest of the star and with the stellar wind is a challenging
problem in radiation-hydrodynamics, but one that is tractable.

One often sees allusions to both a ``hypernova'' (Paczynski 1998)
model and a collapsar model. We make no distinction here. We avoid the term
hypernova as applied to GRBs because one of us previously used the
same word to mean a super-bright pair-instability supernova (Woosley
\& Weaver 1982). However, to the extent that the term hypernova is
used by the GRB community, it is an observational phenomenon caused by
a collapsar.

\section{Supernova Fallback}
\label{fallback}

The simplest way, conceptually, to form a black hole in a massive
star, and thus set up the conditions for the collapsar model, is for
the traditional neutrino powered explosion to fail. The iron core
collapses and within a second or so has made a black hole into which
the rest of the star proceeds to accrete. This may be the common case
for stars above about 35 - 40 M\sun (Fryer 1999), although
uncertainties in convection, mass loss, rotationally induced mixing,
and the explosion mechanism itself make this an uncertain number - and
one that may vary with redshift and metallicity. If the star loses its
hydrogen envelope along the way, and if the jet produced by the
accretion maintains its energy and focus for a longer time than it
takes the jet to tunnel through the star, about 5 - 10 s, a common GRB
is produced (MW99). Otherwise a weaker, less collimated GRB results
(helium star case; MacFadyen \& Woosley 1998), or an energetic
asymmetric supernova ($\S$\ref{diversity}).

\begin{figure}[t]
\psfig{file= 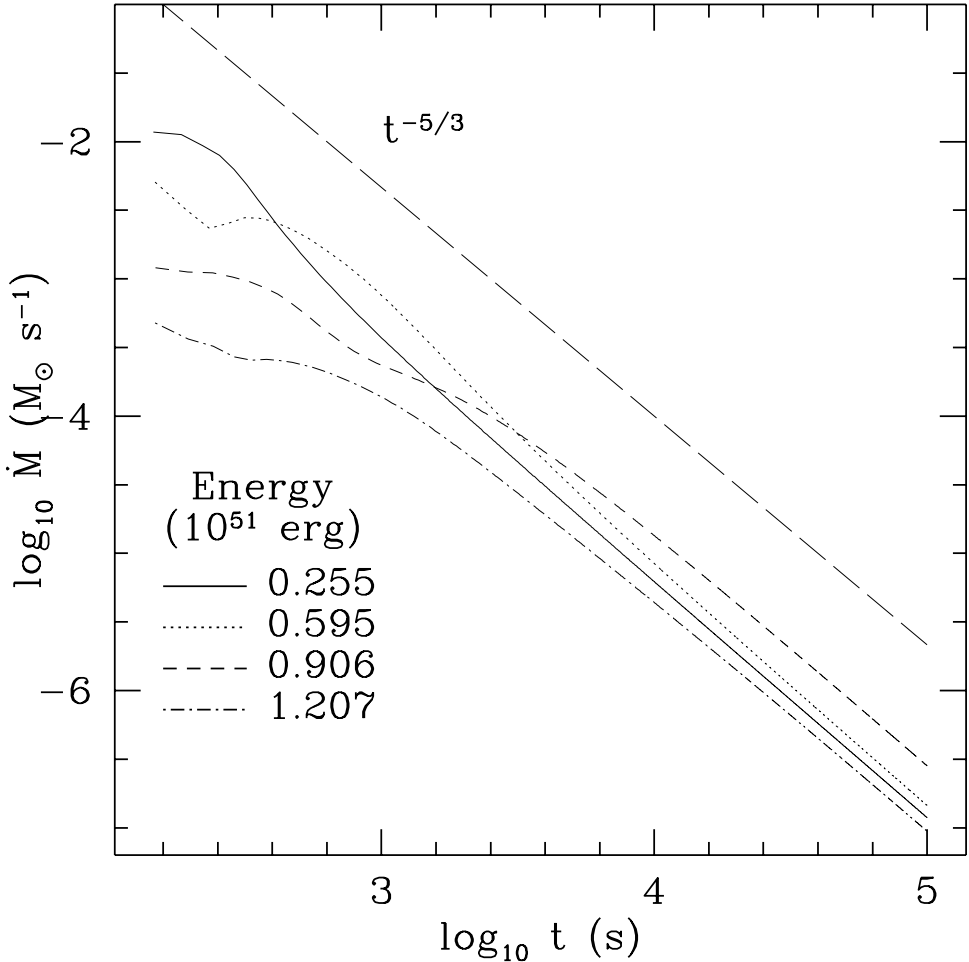,width=\textwidth}
\label{fig1}
\caption{Accretion rates for fall back in five different explosions of
a 25 M\sun \ presupernova star (see text). These five explosions gave
kinetic energies at infinity for their ejecta of 0.255, 0.595, 0.906,
and 1.207 $\times 10^{51}$ erg. The integrated fall back masses for
these spherically symmetric calculations were 3.71, 2.85, 1.39 and
0.48 M\sun \ respectively. Characteristic time scales for the fallback
are 100, 450, 1140, and 1060 s. Calculations were carried out using a
one-dimensional version of the PROMETHEUS hydrodynamics code.}
\end{figure}

However, there should also be a range of stellar masses for which a
black hole is not made promptly, but after a ``successful'' shock has
already been launched. The binding energies of stellar helium cores
outside the collapsing iron core increases with their mass. The
energy of the neutrino engine seems, if anything, to decrease with
mass. Thus there is a range of masses, estimated by Fryer to be
roughly 20 to 40 M\sun, where a supernova occurs, but so much matter
fails to achieve escape and falls back onto the neutron star that it
turns into a black hole. This delayed production of a black hole is
probably a more frequent occurence than prompt black hole formation.

As a representative case, consider a 25 M\sun \ main sequence star
evolved with mass loss, rotationally induced mixing, and angular
momentum transport (Heger, Woosley, \& Langer 1999). This star ends
its life as a red supergiant with an iron core of 1.90 M\sun, a helium
core of 8.06 M\sun, and a low density envelope of 6.57 M\sun \ (total
mass 14.6 M\sun). The presupernova stellar radius is $8.1 \times
10^{13}$ cm.  The model has sufficient angular momentum in the equator
($j \sim 10^{17}$ cm$^2$ s$^{-1}$) to form an accretion disk outside
the black hole.  

Explosions were simulated in this star using a piston at the edge of
the iron core (MacFadyen, Woosley, \& Heger 1999). The motion of this
piston was varied so as to produce a kinetic energy at infinity for
the ejecta ranging from 0.255 $\times 10^{51}$ erg (Model 25A1) to 2.09
$\times 10^{51}$ erg (Model 25A16). The subsequent evolution was
followed using two different one-dimensional hydrodynamics codes,
KEPLER (an implicit Lagrangian hydrodynamics code) and PROMETHEUS (an
explicit Eulerian code).  For similar assumptions
regarding the launch of the shock and the inner boundary condition,
the results of the two codes agreed. For energies above $1.5 \times
10^{51}$ erg, all matter external to the piston was ejected, but for
lower energies an increasing amount of mass fell back to the origin
(Fig. 1). At late times the accretion rate followed the t$^{-5/3}$
scaling predicted by Chevalier (1989).

It is noteworthy that the accretion rate during the time most of the
mass falls back, about 0.001 to 0.01 M\sun \ s$^{-1}$, is very similar
to that frequently invoked in the MHD version of the merging neutron
star model ($\S$\ref{scenarios}), especially for the lower explosion
energies. If jets are to form in one place, surely they should form in
the other. However, for these relatively low accretion rates, the disk
temperature will be too cool to emit neutrinos efficiently. Any jet
that forms must be powered by MHD processes. If we make a simple {\sl
ansatz} that the jet energy, at any point in time, is an efficiency
factor, $\epsilon$, times $\dot M c^2$, with $\epsilon \sim 0.001 -
0.01$ (certainly modest compared to many assumptions in the
literature), then the energy potentially available for making a jet in
Model A01 is $\sim 10^{52} - 10^{53}$ erg. This is large compared both
to the energy of the initial shock in Model 25A1 and the energy of a
typical supernova.

\section{Some General Considerations}
\label{general}

Unfortunately, while a compelling case can be made, both on
observational grounds (e.g., Livio 1999; Pringle 1993) and from theory
(MacFadyen, Woosley, \& Heger 1999) for linking the jet energy to the
accretion rate, the energy alone does not define the model. One
still needs to know the initial partition between internal and kinetic
energy and the beaming angle. In ``thermal'' models, such as the
neutrino version of merging neutron stars or collapsars, the initial
energy is overwhelmingly in the form of radiation and pairs. In fact,
the plasma starts at rest with $aT^4/\rho c^2 \sim \gamma_f \sim
100$. Expansion of the radiation converts internal energy into kinetic
energy very far from the source. For Poynting flux models, on the
other hand, the jet may be born relatively cold.  The initial
collimation of the jet may be either by pressure and density
gradients, as in the collapsar of MW99, or by magnetic fields, or
both. Lacking details of the jet formation process, ambiguity in the
collimation angle and mass to energy ratio makes predictions
difficult, but hot, poorly collimated jets will clearly have a harder
time penetrating the star.

One also expects some systematic differences between cases in which
the black hole forms promptly (Case A) or by fall back (Case B) that
may bear on this issue of collimation.  The lower accretion rate in
Case B suggests a smaller disk mass in steady state (Popham et
al. 1999) and the confining pressure of the medium through which the jet
initially propagates will also be less in Case B, because the star has
already partly exploded. In both Case A and B there will still be an
inner disk that will help to collimate the initial outflow, but,
depending upon how much mass falls back and its angular momentum, that
disk may not extend to such large radii in Case B. All in all, one
expects that the geometrical focusing of the jet at least, may not be
so great in Case B, especially for thermal models. The extent of MHD
collimation is, however, unknown.

Given an initially well collimated jet, one still faces a formidable
computational task following its propagation out to, say, 1000
Schwarzschild radii.  The jet is an inherently relativistic and
can only be described accurately by a special relativistic (SR)
calculation. To do less gives, at best, a qualitative description of
the jet propagation while possibly generating unrealistic artifacts
such as superluminal speeds (MW99). Special relativistic codes are
available (e.g., Aloy et al.1999) and can be adapted to the problem,
but, unfortunately, results are not yet available.

There are several SR effects worth keeping in mind though. First, a
jet of radiation and matter has quite different properties, in SR,
from one composed only of matter. In particular, the equivalent
``dynamical'' density, which must be regarded as a vector, is related
to the rest mass density, $n$, by (Rosen et al. 1999) \begin{equation}
\rho = 2 n \gamma^2 \left( {\gamma \over \gamma + 1} + {\Gamma p \over
(\Gamma - 1) nc^2} \right) \end{equation} which clearly shows the
increase of the effective $\rho$ with $\gamma$ and $p$. Here $\gamma =
(1 - (v/c)^2)^{-1/2}$, $n$ is the rest mass density, $\Gamma$, the
adiabatic index, and $p$, the pressure. As noted earlier, for a
thermal model, $p/n c^2$ is initially about 100. As $p$ turns into
$\gamma$ by expansion, the relativistic correction to the momentum
becomes anisotropic and greatest along the jet. As a result, SR jets
of radiation and matter have much more penetrating power than
Newtonian jets with rest mass density, $n$.

Time dilation also plays an important role. In the frame of the jet,
the star is crossed in a shorter time than in the lab frame. Yet
perpendicular to the jet, motions remain sub-relativistic and clocks run
at similar rates. Thus a SR jet loaded with radiation will diverge, in
the laboratory frame, less than a similar Newtonian jet loaded with
radiation. Indeed, in a Newtonian code, the sound speed and the jet
speed would both be $\sim c$.

Together these effects help to explain why a jet, initially focused by
the geometry of the accretion disk or by the magnetic field near the
hole, but loaded with radiation, might maintain its collimation while
its internal energy is converted into kinetic energy. Eventually, if the
star is not too big, the jet escapes, reaches its asymptotic $\gamma$,
and produces a GRB by running into circumstellar material. 

However, we shall be particularly interested here in another case -
jets that lose their energy before breaking out, share that energy
with the star, and thus become only mildly relativistic.  Our present
calculations are, of necessity, carried out using a Newtonian version
of PROMETHEUS, but we have attempted to capture the flavor of mildly
relativistic jets as they propagate through the helium core and red
giant envelope of an exploding star.  To do so, we picked an inner
boundary radius, 10$^9$ cm, which is computationally expedient (i.e.,
not too small), but still well within the helium core, and at about
the radius where radiation and rest mass might start to become
comparable (see, e.g., Fig. 26 of MW99), especially for MHD models in
which the initial thermal loading of the jet is not so large. Besides
the supernova structure when the jet starts to propagate, there are
three key ingredients to the model, all specified at 10$^9$ cm: 1) the
kinetic energy of the jet as a function of time, given by $\epsilon
\dot M c^2$; 2) the opening angle of the jet, assumed to have a 10
degree half-angle, and 3) the ratio of internal pressure to kinetic
energy, $f_{\rm P}$. This last parameter turns out to be quite
important. If the jet pressure is large compared to the stellar
surroundings in which it propagates, the jet will diverge. If it is
less, the jet may, under some circumstances, be hydrodynamically
focused to a still smaller opening angle. For the calculations we
shall consider, the pressure in the jet is dominantly due to radiation
- though not by a large margin for the smaller values of $f_{\rm P}$.

The principal effect of $f_{\rm P}$ is to increase the tendency of the
jet to diverge. This divergence may, in fact have already occurred
inside 10$^9$ cm. For a relativistic jet, the effective value of
$f_{\rm P}$ would actually be much larger owing to the previously
mentioned modification of the dynamical density and time dilation. In
order to keep our jet velocities on our Newtonian grid below $c$
however, we are compelled to study only $f_{\rm P} \ltaprx 1$.

\section{Some Representative Calculations}
\label{result}

\begin{figure}[t]
\psfig{file=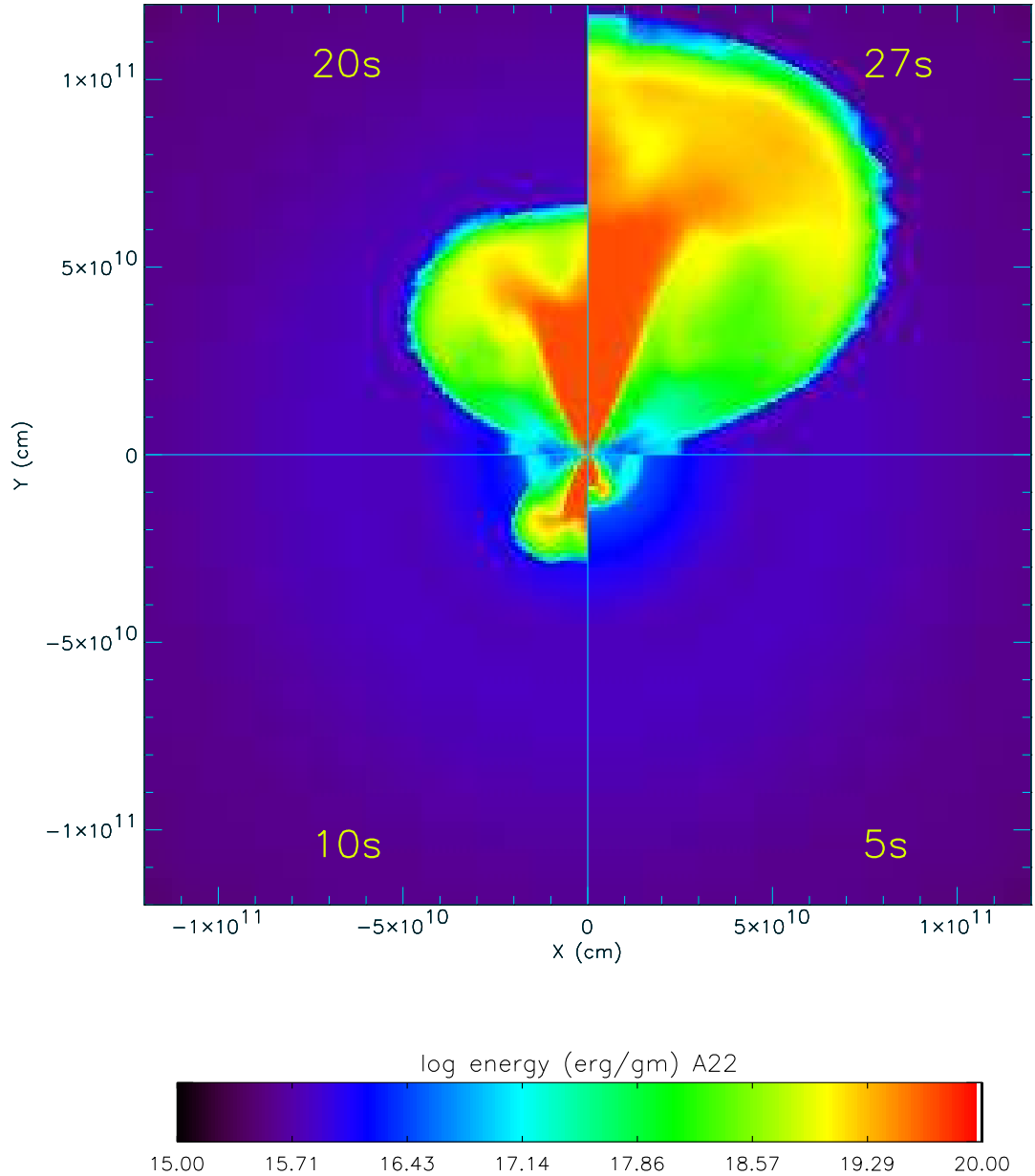,width=\textwidth}
\label{fig2}
\caption{The total energy density of the jet and explosion is shown at
times of 5, 10, 20 and 27 s after initiation for jet Model A22 (Table
1). The passage of the jet initiates a shock that propagates to lower
latitudes, eventually exploding the entire star.  The supernova shock
can be seen at a radius of about $2\times 10^{10}$ cm.}
\end{figure}

To illustrate the possible characteristics of supernovae exploded by
jets, we calculated the two-dimensional evolution of Model 25A1
incorporating parameterized jets.  Details of these and other similar
calculations will be presented in a forthcoming paper (MacFadyen,
Woosley, \& Heger 1999). The spherically symmetric explosion,
followed until 100 s after the launch of a weak shock in the
KEPLER code, was remapped onto the Eulerian grid of a two-dimensional
version of PROMETHEUS. This grid used 150 radial zones spaced
logarithmically between an inner boundary at 10$^9$ cm and the outer
boundary at 8.1$\times$10$^{13}$ cm.  Forty angular zones concentrated
near the pole were used to simulate one quadrant of the stellar
volume, assuming axial and reflection symmetry across the equatorial
plane. The angular resolution varied from 1.25$^\circ$ at the pole to
3.5$^\circ$ at the equator.  At 100 s, the inner 1.99 M\sun \ of the
star was removed and replaced by an open (zero radial gradient of all
variables) boundary condition at 10$^4$ km.  The 1.99 M\sun \
continued to contribute to the gravitational potential as a central
point mass and mass accreting through the inner boundary was added to
the point mass during the calculation. At this time the weak initial
shock was already at 1.1$\times 10^5$ km when the jet was turned on at
the inner boundary.

We gave the jet a constant velocity at this inner boundary, 10$^{10}$
cm s$^{-1}$, a compromise between what the code could realistically
calculate ($v$ less than $c$) and the true relativistic nature of the
initial jet. This velocity, the radius of the inner boundary, and the
(Newtonian) kinetic energy of the jet implied a jet density, $1.9 \times
10^3$ g cm$^{-3}$ $ \ (\dot M/0.01 M\sune \ {\rm s^{-1}})(\epsilon /
0.01)$. This assumed that any internal energy deposited in the jet
near the black hole had been decompressed by adiabatic expansion to
the point where, at 10$^9$ cm, it was small compared to $\rho v^2$.

\begin{figure}[t]
\psfig{file=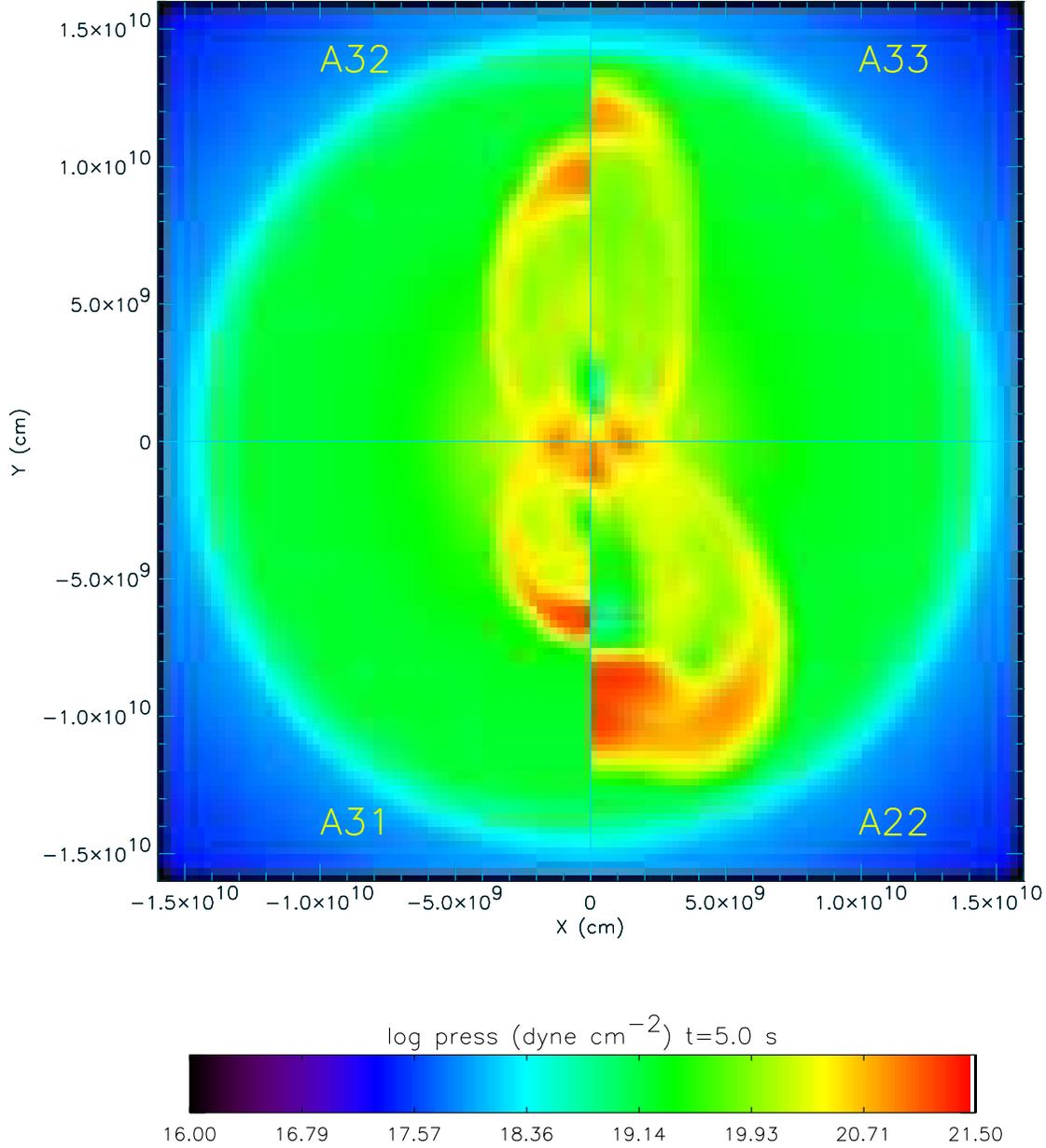,width=\textwidth}
\label{fig3}
\caption{Pressure in the jet and surrounding star at 5.0 s after the 
initiation of the jet in four different models. Higher presure leads
to greater jet divergence, more mass swept up, and slower
propagation. Model A22 had a higher jet energy than the other models
(Table 1).}
\end{figure}

We considered four cases, $\epsilon = 0.001$, $f_{\rm P}$ = 0.001,
0.01, and 0.1 and $\epsilon = 0.01$, $f_{\rm P}$ = 0.01.  The results
are summarized in Table 1 and Figs. 2 - 4. Here the name follows the
convention ``AMN'' where ``A'' indicates the model was based upon the
weakest explosion considered of a 25 M\sun \ (main sequence mass)
supernova (0.255 $\times 10^{51}$ erg; Fig 1), ``M'' is the exponent
of the efficiency factor, $\epsilon = 10^{\rm -M}$, and ``N'' is the
exponent of the pressure factor, $f_{\rm P} = 10^{\rm -N}$.  The mass
accreted, $\Delta M$ in Table 1, is smaller than the 3.71 M\sun \
computed without a jet (Fig. 1) for Model 25A1, because the jet
impeded the accretion at high latitude and because the accretion was
not quite over at after 500 seconds (Fig. 1). The total energy input
by the jet was still $\epsilon \Delta M c^2$, but the number in Table 1
was reduced by the work done up to 500 s in unbinding the star and by
the internal and kinetic energy which passed inside the inner
boundary.  The $2.55 \times 10^{50}$ erg due to the initial shock has
been subtracted in Table 1 so that E$_{tot}$ reflects only the energy
input by the jet.

The angular factor R($\theta > 10^\circ$) is the ratio of the integral
of the kinetic energy due to the jet outside 10 degrees polar angle
(98.5\% of the sky) to the total kinetic energy in the star due to the
jet (see Fig. 4).  These energies were computed by taking the total
kinetic energy at 400 s after jet initiation in both regions and
subtracting the kinetic energy of the initial supernova shock.
R($\theta > 10^\circ$) measures the extent to which the jet spread
laterally and shared its energy with the rest of the star.  The
limiting case R=0 would correspond to a jet that shared none of its
energy with the supernova outside an initial 10$^\circ$ polar angle.
This sort of behavior is expected for ``cold'' jets with internal
pressure small compared to the exploding helium core.  The other
extreme, where the jet shared its energy evenly with the entire star
and produced a spherical explosion, would correspond to R = $\cos
\theta$ = 0.985.  Our ``hot'' jets lie somewhere between these two
limits. The quantity R($\theta > 20^\circ$) was similarly computed for
a polar angle of 20$^\circ$. The isotropic limit there would be 0.940.

\begin{table*}
\vskip 8pt \centerline {TABLE 1: Explosion Characteristics at t = 400
s After Jet Initiation}
\begin{center}
\vskip 3pt
\begin{tabular}{ccccccc}
\hline 
      Name & $\epsilon$  & f$_{\rm P}$   & $\Delta$ M & E$_{tot}$ 
        & R($\theta > 10^\circ$) & R($\theta > 20^\circ$) \\[3pt] 
           &             &               & (M\sun) & (10$^{51}$\,ergs) & & \\
\hline
      A33 &  0.001  & 0.001 & 2.76 & 3.38 & 0.075 & 0.037 \\
      A32 &  0.001  & 0.01  & 2.69 & 3.23 & 0.102 & 0.047 \\
      A31 &  0.001  & 0.1   & 2.51 & 3.00 & 0.425 & 0.256 \\
      A22 &  0.01   & 0.01  & 1.72 & 19.91 & 0.429 & 0.230 \\
\hline
\end{tabular}
\end{center} 
\end{table*} 

In all cases a very energetic asymmetric supernova resulted. Since the
integrated mass of the (Newtonian) jet in our code was comparable to
that of the stellar material within 10 degrees, the time for jet break
out was approximately the stellar radius divided by the jet input
speed. In reality, that would be $\sim R/c$, or for a red supergiant
several thousand seconds. Since the energy of the jet engine had
declined greatly by that time, due to the declining accretion rate
(Fig. 1), and the jet had swept up far more than $\gamma^{-1}$ of its
rest mass, the jet that broke out was only mildly
relativistic. Both the long time scale and the low energy input are 
inconsistent with what is seen in common GRBs. However, if the
hydrogen envelope had been lost, a longer than typical GRB could have
resulted.

\begin{figure}[t]
\psfig{file=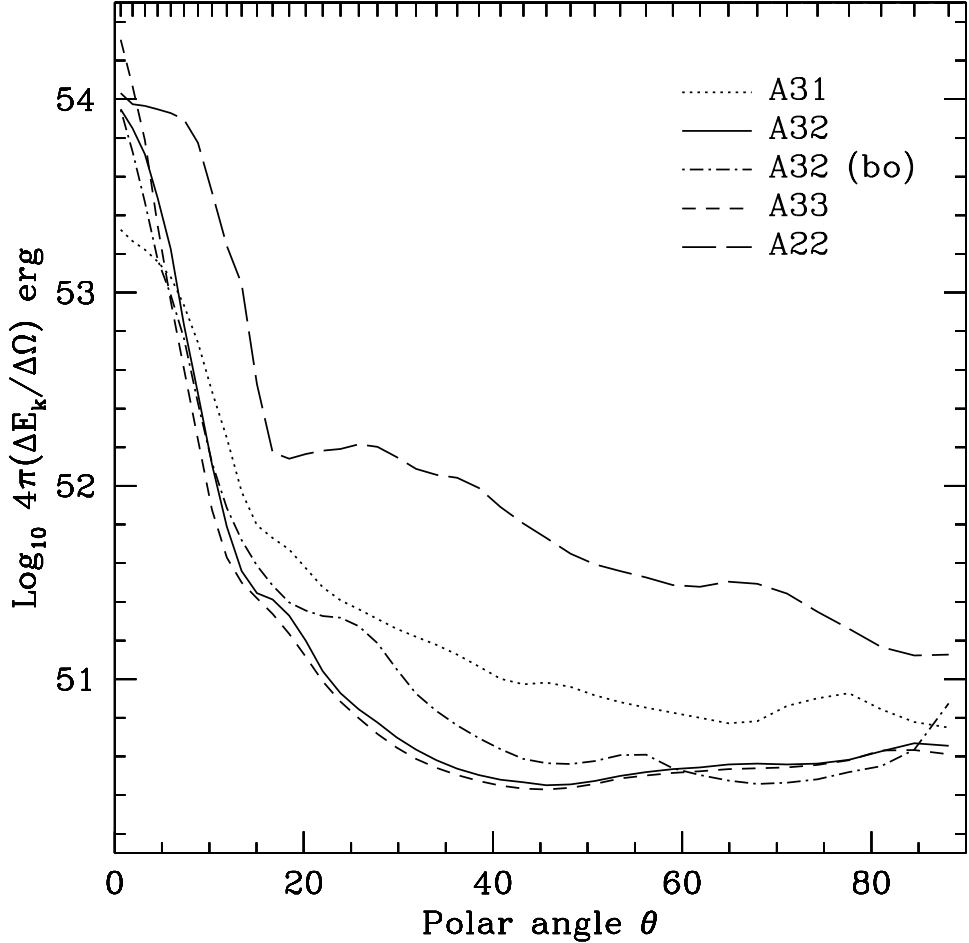,width=\textwidth}
\label{fig4}
\caption{The ``equivalent isotropic kinetic energy'' as a function of
polar angle for four models having variable energy efficiency factors
and internal pressures (Table 1 and text). Model A32 is shown at two
times, once at 400 s after the initiation of the jet and later, at
7716 s, as the jet penetrated the surface of the star at $8 \times
10^{13}$cm; dash-dot line. Other models are also shown for comparison
at 400 s. Note that the degree of collimation is strongly dependent
upon f$_{\rm P}$.  Equivalent isotropic kinetic energy is defined as
the integral from the center to surface of the star of its kinetic
energy in the solid angle subtended by $\theta$ and $\theta +
\Delta\theta$ divided by the solid angle, 2$\pi(\cos\theta -
\cos(\theta + \Delta\theta))$ and multiplied by 4$\pi$.  The injected
energy at the base of the jet would be a flat line out to ten degrees
with a value equal to $66 \, \epsilon \Delta M c^2$ with $\Delta M$ in
Table 1 and 66 = (1 - $\cos(10^\circ))^{-1}$.  Tick marks along the
top axis give the angular zoning of the two dimensional code. }
\end{figure}

Figs. 3 and 4 illustrate how the pressure balance between the jet and
the star through which it propagates affected its collimation
properties. The interaction at late times with the hydrogen envelope
had relatively little effect on the angular energy distribution which
was set chiefly by $f_{\rm P}$ and the interaction with the helium
core.  Model A33 had the lowest internal pressure (note that the
actual value of the initial pressure depends upon the product of
$\epsilon$ and $f_{\rm P}$). The final jet was collimated even more
tightly than given by its initial injection. That is, a jet initially
of 10 degrees half width will exit the star with a FWHM of less than
two degrees, about 0.06\% of the sky (though the angular resolution of
the code is questionable for such small angles). Meanwhile the energy
at larger angles was not much greater than that given by the initial,
weak spherically symmetric explosion, 10$^{50.4}$ erg. There was
little sharing of the jet energy with the star and, except for the
jet, the supernova energy remained low.

This is to be contrasted with Models A22 and A31 where the jet
collimation was much weaker and much more energy was shared with the
star. Note that though Model A22 had about 6 times the total energy of
A31 owing to its larger $\epsilon$ the fraction of energy at large
angles in both these models was significantly greater than in Models
A32 and A33.  Model A22 would be an especially powerful supernova as
well as one accompanied by a jet.

\section{Supernovae and GRB Diversity}
\label{diversity}

Provided the necessary conditions for the collapsar model can be met -
black hole formation in a massive star with sufficient angular
momentum to make a disk - the discussion and results of the previous
two sections suggest a wide variety of possible outcomes, including,
besides ordinary GRBs:

\vskip 0.2 in
\noindent
{\sl``Smothered'' and broadly beamed gamma-ray bursts; GRB 980425} -
These can occur in helium stars in which the jet either fails to
maintain sufficient focus (e.g., is too ``hot'' compared to the star
through which it propagates), or loses its energy input before
breaking out of the star ($\ltaprx$10 s; MW99). An energetic supernova still
occurs (SN 1998bw, in this case) and a weak GRB is produced, not by the
jet itself, but by a strong, mildly relativistic shock from break out
interacting with the stellar wind.  (Woosley, Eastman, \& Schmidt
1999).  Because these events are so low in gamma-ray energy, many could
go undetected by BATSE. Indeed these could be the most common form of
GRB in the universe. Because the initial jet may be less effectively 
collimated in GRBs made by supernova fall back, it is tempting to
associate these phenomena with delayed black hole formation and the
stronger GRBs with prompt black hole formation. More study is needed.

\vskip 0.2 in
\noindent
{\sl Long gamma-ray bursts; $\tau_{\rm burst} \gtaprx 100$ s} - Though
typical ``long, complex bursts'' observed by BATSE last about 20
seconds, there are occasionally much longer bursts. For example,
GRB950509, GRB960621, GRB961029, GRB971207, and GRB980703 all lasted
over 300 s. These long durations may simply reflect the light crossing
time of the region where the jet dissipates its energy (modulo
$\gamma^{-2}$), especially in the ``exterior shock model'' for
GRBs. However, if the event is due to internal shocks, the duration
depends on the time the engine operates. Such long bursts would imply
enduring accretion on a much longer time scale than one expects in the
simplest collapsar model where the black hole forms promptly. The
fallback powered models discussed in this paper could maintain a GRB
for these long time scales (Fig. 1).

\vskip 0.2 in
\noindent
{\sl Very energetic supernovae - SN 1997cy} - Germany et al. (1999)
have called attention to this extremely bright supernova with an
unusual spectrum. The supernova was Type IIn and its late-time light
curve, which approximately followed the decay rate of $^{56}$Co, would
require $\gtaprx$2 M\sun \ of $^{56}$Ni to explain its
brightness. Perhaps this was a pair-instability supernova (Woosley \&
Weaver 1982; Heger, Woosley, \& Waters 1999). On the other hand,
circumstellar interaction could be the source of the energy and the
agreement with $\tau_{1/2}$($^{56}$Co) merely fortuitous. This would
require both a very high explosion energy and a lot of mass loss just
prior to the supernova. The sort of model described in
$\S$\ref{result}, especially Model A22, could provide the large energy
in a massive star that would be naturally losing mass at a high rate
when it died. But the radius is too large and the jet would share its
energy with too great a mass to make a common GRB. Therefore we regard
the detection of a short, hard GRB from the location of SN 1997cy as
spurious.

\vskip 0.2 in
\noindent
{\sl Nucleosynthesis - $^{56}$Ni and the $r$-process} - An explosion
of 10$^{52}$ erg focused into 1\% of the star (or 10$^{53}$ erg into
10\%) will have approximately the same shock temperature as a function
of radius as an isotropic explosion of 10$^{54}$ erg. From the simple
expression ${{4} \over {3}} \pi r^3 a T^4 \sim 10^{54}$ erg (Woosley
\& Weaver 1995), we estimate that a shock temperature in excess of 5
billion K will be reached for radii inside $4 \times 10^9$ cm. The
mass inside that radius external to the black hole (assumed mass
initially 2 M\sun) depends on how much expansion (or collapse) the
star has already experienced when the jet arrives. Provided the star
has not expanded much before the jet arrives, an approximate number
comes from the presupernova model, 3 M\sun \ times the solid angle of
the explosion divided by 4 $\pi$, or $\sim$0.03 M\sun.  Additional
$^{56}$Ni is probably synthesized by the wind blowing off the
accretion disk (MW99; Stone, Pringle, \& Begelman 1999) and this may
be the dominant source in supernovae like SN 1998bw.

The composition of the jet itself depends upon details of its
acceleration that are hard to calculate. However it should originate
from a region of high density and temperature (Popham, Woosley, \&
Fryer 1999). The high density will promote electron capture and lower
$Y_e$. The high entropy, low $Y_e$, and rapid expansion rate are what
is needed for the $r$-process (Hoffman, Woosley, \& Qian 1997). The
mass of the jet, $\sim10^{-4}$ M\sun \ (corrected for relativity) is
enough to contribute significantly to the $r$-process in the Galaxy
even if the event rate was $\ltaprx$1\% that of supernovae and the jet
carried only a fraction of its mass as $r$-process.

\vskip 0.2 in
\noindent
{\sl Soft x-ray transients from shock breakout} - Focusing a jet of
order 10$^{52}$\,ergs into 1 - 10\% of the solid angle of a supernova
results in a shock wave of extraordinary energy (Fig. 4).  As it nears
the surface of the star, this shock is further accelerated by the
declining density gradient. MacFadyen, Woosley, \& Heger (1999)
estimate, for a 10$^{54}$ erg (isotropic equivalent) shock, a break
out transient of 10$^{49}$ erg s$^{-1}$ (times $(1-\cos\theta_j)$, the
solid angle of the jet at break out divided by 4$\pi$, where $\theta_j$
is the half opening angle of the jet at breakout) for $\sim$10 s. The
color temperature at peak would be approximately $2 \times 10^6$ K
(see also Matzner \& McKee 1999). A 10$^{53}$ erg shock gave a
transient about half as hot and ten times longer and fainter. The
impact of the mildly relativistic matter could give an enduring x-ray
transient like the afterglows associated with some GRBs, even though
the time scale is too long for the x-ray burst to be a common GRB
itself.

\vskip 0.2 in
\noindent
{\sl Mixing in supernovae - SN 1987A} - It is generally agreed (Arnett
et al. 1989) that the explosion that gave rise to SN 1987A initially
produced a neutron star of approximately 1.4 M\sun.  There may have
been $\sim$0.1 M\sun \ of fallback onto that neutron star (Woosley
1988) and a black hole may or may not have formed. Again invoking our
{\sl ansatz} that ${\rm L_{jet}} = \epsilon {\rm \dot M} c^2$, even
for $\epsilon \sim$ 0.003, we have a total jet energy of $6 \times
10^{50}$ erg. This is about half of the total kinetic energy inferred
for SN 1987A. Thus very appreciable mixing and asymmetry would be
introduced by such a jet - {\sl provided the material that fell back
had sufficient angular momentum to accumulate in a disk outside the
compact object}. However this would not be enough energy to make a
powerful gamma-ray burst as proposed by Cen (1999).

\vskip 0.2 in
\noindent
{\sl Still to be discovered} - It may be that, especially with common
GRBs, we have just seen the ``tip of the iceberg'' of a large range of
high energy phenomena powered by hyper-accreting, stellar mass black
holes. We already mentioned the possibility of a large population of
faint, soft bursts like GRB 980425. Other possibilities include very
long GRBs below the threshold of BATSE, ``orphan'' x-ray 
afterglows from jet powered Type II supernovae, supernova remnants
having toroidal structure, GRBs from the first explosions of 
massive stars after recombination, and more. It is an exciting time. 

\section{Does It All Work?}

Exciting that is, if it all works as described.  That a
hyper-accreting black hole (M$_{\rm hole}$ = 2 to 10 M\sun, accreting
10$^{-1}$ - 10$^1$ M\sun \ s$^{-1}$) gives rise to an energetic jet
with dramatic observational consequences seems to us unavoidable. True
the physics of jet formation is poorly understood, but the ubiquity of
jets in all sorts of systems where disk accretion is going on, the
success of the basic idea of AGN's as accreting massive black holes,
and the identity of ``microquasars'' as accreting black holes all
argue that this is an assumption worth exploring. That supernovae
sometimes form black holes, both promptly and in a delayed manner,
also seems unavoidable. Our calculatiions show that if a jet forms in
a massive collapsing star, and if that jet has only a fraction of a
per cent of the energy potentially available from the accretion
process, that energetic supernovae and GRBs are a likely outcome.

The weakest assumption in all the models discussed here is that the
requisite amount of angular momentum is present to form a disk. The
best available stellar evolution models suggest it is there (Heger,
Langer, \& Woosley 1999), but these calculations have left out
magnetic field effects that might lead to the dramatic slowing of the
rotation of the helium core, especially in red supergiants (Spruit \&
Phinney 1998). These models also imply that neutron stars may be born
rotating near break up. Whether either of these concerns will
ultimately prove fatal to the model remains to be seen. Since GRBs are
much rarer in the universe than supernovae, it is of
course possible that the production of GRBs demands some very special
circumstances, e.g., the merging of two stripped helium stars already
in a late stage of evolution (Fryer, Woosley, \& Hartmann 1999). 

\section*{Acknowledgments}

This work has been supported by NASA (NAG5-8128 and MIT SC A292701),
the NSF (AST 97-31569), and the Department of Energy ASCI Program
(W-7405-ENG-48), and by the A. V. Humboldt-Stiftung (1065004).

\end{document}